\documentclass[aps,prb,twocolumn,superscriptaddress,longbibliography]{revtex4-2}

\usepackage{amsmath}
\usepackage{amssymb}
\usepackage{graphicx}
\usepackage{bm}
\usepackage{color}
\usepackage{hyperref}
\usepackage{booktabs}
\usepackage{subcaption}

\newcommand{\kx}{k_x}
\newcommand{\ky}{k_y}
\newcommand{\coskx}{\cos k_x}
\newcommand{\cosky}{\cos k_y}
\newcommand{\dx}{d_{x^2-y^2}}
\newcommand{\dz}{d_{z^2}}
\newcommand{\Dp}{\Delta_\perp}
\newcommand{\Ds}{\Delta_s}
\newcommand{\Da}{\Delta_\alpha}
\newcommand{\Db}{\Delta_\beta}
\newcommand{\Dg}{\Delta_\gamma}
\newcommand{\tperp}{t^\perp}
\newcommand{\Jxz}{J_{xz}}
\newcommand{\Jperp}{J_\perp}
\newcommand{\Tc}{T_c}

\newcommand{\Dxz}{\Delta_{xz}}

\newcommand{\spm}{s^\pm}

\begin{document}

\title{Pairing Mechanism  in Bilayer Nickelate La$_3$Ni$_2$O$_7$ Superconductors}

\author{Xianxin Wu}
\affiliation{Institute for Theoretical Physics, Chinese Academy of Sciences, Beijing, China}
\author{Tao Xiang}
\affiliation{Beijing National Laboratory for Condensed Matter Physics,
Institute of Physics, Chinese Academy of Sciences, Beijing 100190, China}
\affiliation{School of Physical Sciences, University of Chinese Academy of Sciences, Beijing 100049, China}
\author{Jiangping Hu}\email{jphu@iphy.ac.cn}
\affiliation{Beijing National Laboratory for Condensed Matter Physics,
Institute of Physics, Chinese Academy of Sciences, Beijing 100190, China}
\affiliation{School of Physical Sciences, University of Chinese Academy of Sciences, Beijing 100049, China}
\affiliation{New Cornerstone Science Laboratory,
Institute of Physics, Chinese Academy of Sciences, Beijing 100190, China}

\date{\today}

\begin{abstract}
The recent discovery of superconductivity with $T_c \approx 80$~K in bilayer nickelate La$_3$Ni$_2$O$_7$ provides a new setting in which to test the organizing principles of unconventional high-temperature superconductivity. We show that the gene principle and the collaborative Fermi-surface rule which were previously proposed to unify unconventional high temperature superconductors, extend naturally to this bilayer, multi-orbital system. We identify that there are two antiferromagnetic exchange channels that can provide the dominant pairing force: an interlayer intra-orbital nearest-neighbour exchange $J_\perp$ between $d_{z^2}$ orbitals mediated by the inner apical oxygen, and an intralayer inter-orbital nearest-neighbour exchange $J_{xz}$ between $d_{z^2}$ and $d_{x^2-y^2}$ orbitals mediated by the in-plane oxygen. Owing to the bilayer bonding--antibonding splitting and the $B_{1g}$ symmetry of the $d_{x^2-y^2}$ orbital, these two channels cooperate to produce a  robust  $s^\pm$ superconducting state with an internal sign reversal between mirror-even and mirror-odd Fermi-surface pockets in momentum space.  Both pairing channels maximize the superconducting gap on the $\beta$ pocket  with a form factor $(cosk_x-cosk_y)^2$ in momentum space. The result places La$_3$Ni$_2$O$_7$ within a unified framework for unconventional superconductivity while revealing a distinct electronic environment for high-$T_c$ pairing.
\end{abstract}

\maketitle

\section{Introduction}

The discovery of high-temperature superconductivity in cuprates~\cite{Bednorz1986} and iron-based superconductors~\cite{iron-based_2008}  have called  universal organizing principles of strongly correlated electrons to understand unconventional superconductivity.  More recently, superconductivity with $T_c \approx 80$~K reported in the bilayer nickelate La$_3$Ni$_2$O$_7$ (Ni327) under pressure~\cite{Sun2023} has provided a new arena in which to test those principles.

In earlier work~\cite{Hu2015prx,Hu2016,Hu2012}, two related ideas were proposed to unify the cuprates and iron-based superconductors. The first is the \emph{gene principle}\cite{Hu2015prx,Hu2016}, which states that unconventional high-$T_c$ superconductivity requires a quasi-two-dimensional electronic environment in which transition-metal $d$ orbitals, strongly hybridized with anion $p$ orbitals, are isolated near the Fermi energy. In such a setting, superconductivity is driven primarily by antiferromagnetic (AFM) superexchange mediated through anions, whereas direct cation--cation exchange can not overcome   repulsive nature between two electrons and disfavors pairing. The second is the \emph{collaborative Fermi-surface rule}~\cite{Hu2012}, which selects the pairing symmetry by maximizing the overlap between the pairing form factor generated by AFM exchange and the actual Fermi surface.

These two ideas account naturally for both cuprates and iron-based superconductors. In cuprates, nearest-neighbour (NN) Cu--O--Cu superexchange $J_1$ selects the d-wave form factor $(\coskx-\cosky)$, which overlaps optimally with the large hole Fermi surface~\cite{Scalapino1995}. In iron-based superconductors, next-nearest-neighbour (NNN) Fe--As/Se--Fe superexchange $J_2$ favours an extended s-wave form $(\coskx\cosky)$~\cite{Hu2012,Hu2016}. The latter state is a sign-reversing superconductor in momentum space when both electron and hole pockets coexists in the FeAs families. The two families therefore represent distinct symmetry realizations of the same underlying mechanism on the square lattice.

It can be easily noticed that Ni327 appears to satisfy the gene principle: a quasi-two-dimensional NiO$_2$ environment in which both Ni $e_g$ orbitals strongly hybridize with oxygen and are isolated to dominate the low-energy states near Fermi energy.  What distinguishes Ni327 is not the basic mechanism, but the structure of the problem. First, it is intrinsically a \emph{bilayer} system, with two NiO$_2$ planes per unit cell coupled through apical oxygens. Second, two distinct $e_g$ orbitals, $d_{z^2}$ and $d_{x^2-y^2}$, remain active near the Fermi level, unlike the single-orbital situation in cuprates. Third, the large interlayer hopping $t^{\perp}_{zz}$ between $d_{z^2}$ orbitals through the inner apical oxygen~\cite{YaoDX} generates strong bonding--antibonding splitting and qualitatively reshapes the Fermi surface. Together, these features make the pairing problem intrinsically bilayer and multi-orbital. So far, the pairing mechanism of Ni327 has been under intensive theoretical study, and a consensus is yet to be reached~\cite{Wang327prb,Lechermann2023,Hirofumi2023possible,XWu,ZhangGM_cpl2023, lu2024interlayer,HYZhangtype2,FangYang327prl,WeiLi327prl,YifengYang327prb,YifengYang327prb2,YiZhuangYouSMG,tian2023correlation,Dagotto327prb,zhang2024structural,Jiang_2024,liao2023electron,ryee2024quenched,luo2023hightc,fan2023superconductivity,KuWeiprl,zhan2024cooperation,ChenHH2025,PhysRevB.111.144514,GuanGJ2025}.

\begin{figure}[h!]
    \centering
     \includegraphics[width=0.6\linewidth]{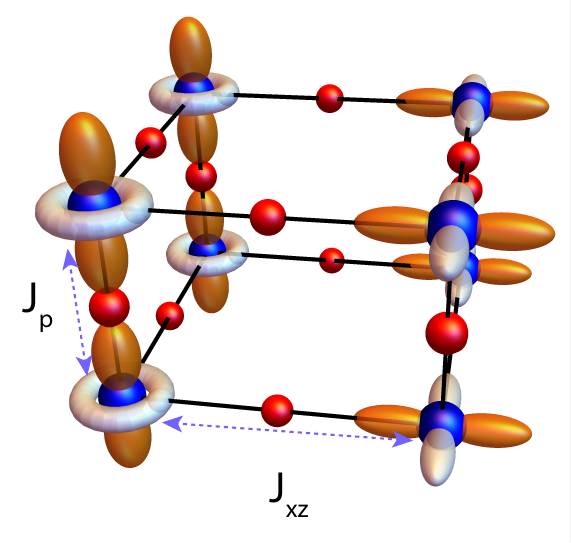}
        \caption{The local AFM exchange couplings: the interlayer AFM exchange between $\dz$ orbitals and the intralayer inter-orbital AFM exchange between $\dz$and $\dx$ orbitals.}
\label{fig1}
\end{figure}
\begin{figure}[h!]
    \centering
     \includegraphics[width=0.6\linewidth]{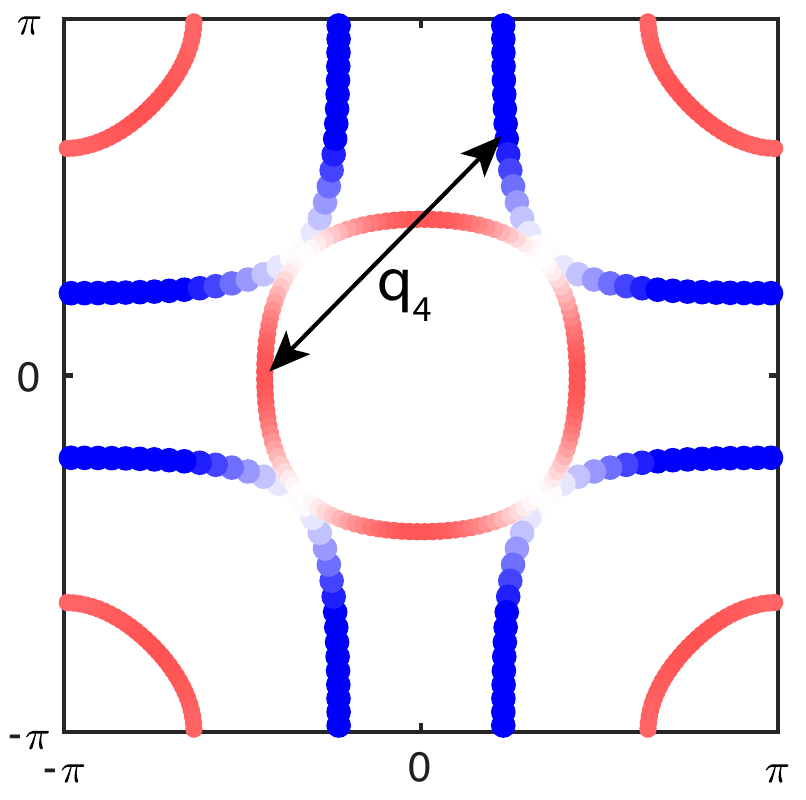}
        \caption{The $s^\pm$ state:  the sign distribution of the superconducting order parameters on Fermi surfaces.  The red color and the blue color represent opposite sign.}
\label{fig2}
\end{figure}

The central question is therefore whether the collaborative Fermi-surface rule can survive in this more complex setting. In this work, we show that it does. We identify two dominant AFM exchange interactions in Ni327: an interlayer NN exchange $J_\perp$ between $d_{z^2}$ orbitals mediated by the inner apical oxygen, and an intralayer NN inter-orbital exchange $J_{xz}$ between $d_{z^2}$ and $d_{x^2-y^2}$ orbitals mediated by in-plane oxygen as illustrated in Fig.\ref{fig1}. Because the $d_{x^2-y^2}$ orbital has $B_{1g}$ symmetry, the inter-orbital pairing channel acquires a non-trivial momentum structure. Once projected to the band basis, the interlayer and intralayer channels cooperate rather than compete, and the resulting gap function achieves its largest overlap on the strongly hybridized $\beta$ pocket. The two pairing channels result in an  $s^\pm$  state  with an internal sign reversal in momentum space as shown in Fig.\ref{fig2}.   There is a sign reversal between mirror-even and mirror-odd Fermi-surface pockets generated by the bilayer structure. Ni327 therefore realizes a unique  superconducting state to test the pairing mechanism for unconventional high temperature superconductivity.  The main results are illustrated in Fig.\ref{fig1}.

\section{Electronic Structure and Fermi Surface Classification}
The electronic structure in the high-pressure tetragonal ($I4/mmm$) phase of Ni327 has been derived~\cite{Sun2023,YaoDX,YZhang2023,Lechermann2023,Hirofumi2023possible,XWu,XJZhou2023,HHWen2023}. Each Ni atom sits in a NiO$_6$
octahedron with average valence Ni$^{2.5+}$, distributing $n = 1.5$ electrons across
the $e_g$ orbitals ($\dx$ and $\dz$) per site.  There are three leading hopping parameters mediated by oxygens in the effective model for the electronic structure.  The inner apical oxygen mediates a
large interlayer hopping $\tperp_{zz}$ between the $\dz$ lobes
pointing along the c-axis. The in-plane oxygens mediate NN intralayer hoppings:
$t_x$ for $\dx$--$\dx$ and inter-orbital hopping $t_{xz}$ for $\dz$--$\dx$. 

We can use the bilayer mirror symmetry $\mathcal{M}_z$, which relates the top and bottom layers,  to classify the electronic bands into layer-bonding (L$+$, mirror-even, eigenvalue $+1$)
and layer-antibonding (L$-$, mirror-odd, eigenvalue $-1$) sectors. Moreover, due to the relatively large $d$-wave inter-orbital hopping $t_{xz}$, the electronic bands can also be classified independently by the relative phase between $d_{x^2-y^2}$ and $d_{z^2}$ orbitals along nearest-neighbor bonds. Specifically, the two orbitals can be combined such that their overlapping in-plane lobes possess either the same sign (O$+$, in-phase) or opposite signs (O$-$, out-of-phase).
 There are three important Fermi pockets: the \textbf{$\alpha$}  pocket centered at $\Gamma$, the large \textbf{$\beta$} pocket resembles the Fermi surface of cuprates, and  the small \textbf{$\gamma$}  pocket near $M$.  Under the mirror symmetry, the $\alpha$  and  $\gamma$ pocket  are mirror symmetric, namely L$+$, while  the $\beta$ is mirror antisymmetric, namely L$-$. In terms of relative phase between two orbitals, a
crucial observation is that $\beta$ and $\gamma$ pockets share the out-of-phase character (O$-$), while the $\alpha$ pocket is of in-phase character (O$+$). The complete pocket classification is given in Table~\ref{tab:pockets}.  
It is important to note that the $\beta$ pocket exhibits strong orbital hybridization.

\begin{table}[h]
\centering
\caption{Classification of Fermi surface pockets based on layer mirror symmetry and
relative phase between two orbitals. The mirror eigenvalue refers to $\mathcal{M}_z$ (bilayer reflection).}
\label{tab:pockets}
\begin{ruledtabular}
\begin{tabular}{ccccc}
Pocket & Mirror & Orbital & Dominant & Location \\
       & eigenvalue & & character & \\
\hline
$\alpha$ & $+1$ (even) & O$+$ (in-phase) & $\dx$ & $\Gamma$ \\
$\gamma$ & $+1$ (even) & O$-$ (out-of-phase) & $\dz$ & $M$  \\
$\beta$  & $-1$ (odd)  & O$-$ (out-of-phase) & $\dx,\dz$  & $M$ \\
\end{tabular}
\end{ruledtabular}
\end{table}

\section{Two Leading AFM Exchange Interactions}

We seek the leading AFM exchange interaction in the strong electron-electron correlation limit.  Before the derivation, it is important to address the filling on orbitals.   The orbital filling distribution is highly nonuniform. The $\dz$ bonding state
($\gamma$ pocket) is near half-filling---the strongly correlated sector. The
$\dx$ orbital contributes approximately $1/4$ filling across $\alpha$ and $\beta$,
far from half-filling. Consequently, the NN $\dx$--O--$\dx$ superexchange responsible
for $d$-wave pairing in cuprates is strongly suppressed in Ni327. The dominant
magnetic exchange must instead involve the near-half-filled $\dz$ orbital, through
both the interlayer and intralayer pathways analyzed in the next section.

The dominant interlayer exchange is the $\dz$--$\dz$ superexchange through the
inner apical oxygen, generated by the large interlayer hopping $\tperp_{zz}$:
\begin{equation}
H_{\perp} = \Jperp \sum_{\langle i\rangle} \mathbf{S}_{i_1,z}\cdot\mathbf{S}_{i_2,z},
\qquad \Jperp = \frac{4(\tperp_{zz})^2}{U},
\label{eq:Jxz}
\end{equation}
where $U$ is the onsite Hubbard interaction between the same orbitals.  This is an AFM exchange between $\dz$ spins on opposite layers on the same Ni site, generated through the apical oxygens between the two NiO$_2$
layers.


The second exchange interaction is  the intralayer inter-orbital AFM exchange, which is mediated  by the inter-orbital NN hopping $t_{xz}$.  In strong correlation limit, at the second order in perturbation theory, an intralayer inter-orbital AFM
exchange:
\begin{equation}
H_J = \Jxz \sum_{\langle ij,\alpha\rangle} \mathbf{S}_{i_\alpha,z}\cdot\mathbf{S}_{j_\alpha,x},
\qquad \Jxz = \frac{2t_{xz}^2}{U+\Dxz }+\frac{2t_{xz}^2}{U-\Dxz},
\label{eq:Jxz}
\end{equation}
where $\Dxz = \varepsilon_x - \varepsilon_z > 0$ is the crystal field splitting.
This is an AFM exchange between the localized $\dz$ spin on
site $i$ and the $\dx$ spin on NN site $j$, mediated by the in-plane oxygen.  
The two exchanges $\Jperp$ and $\Jxz$ are generated by fundamentally different
anion-mediated pathways. $\Jperp$ is on vertical c-axis bond between same orbitals ($\dz$--$\dz$), through the $p_z$ orbital of the apical O.
 $\Jxz$ is on  horizontal ab-plane NN bond between different orbitals ($\dz$--$\dx$), through the $p_{x,y}$ orbital of the in-plane O.

\section{Pairing Symmetry: The $s^\pm$ State}

The interlayer exchange $\Jperp$ acts between two $\dz$ orbitals on opposite
layers.  Decoupling $\Jperp$ in the particle-particle channel accordingly generates only
an \textit{s-wave} interlayer singlet pairing, namely, $\langle d_{z^2\uparrow,1}(\bm{k}) d_{z^2\downarrow,2}(-\bm{k})-d_{z^2\downarrow,1}(\bm{k})d_{z^2\uparrow,2}(-\bm{k})\rangle=\Dp$. The key consequence of this s-wave interlayer pairing is a sign change between
Fermi-surface pockets of even and odd mirror symmetry under $M_z$. Since $\Jperp$
couples the two layers antiferromagnetically, the resulting interlayer Cooper pairs
have even (bonding) and odd (antibonding) combinations across the bilayer. In the
band basis, the mirror-even pockets $\alpha$ and $\gamma$ (L$+$, eigenvalue $+1$)
acquire gap $+\Dp$, while the mirror-odd pocket $\beta$ (L$-$, eigenvalue $-1$)
acquires gap $-\Dp$. This sign change is the direct bilayer analogue of the
$\spm$ sign reversal between $\Gamma$ and $M$ pockets in iron pnictides, here
encoded in the mirror quantum number of the bilayer rather than in the in-plane
Fermi-surface topology.  It is important to notice that this  interlayer exchange would not contribute to  pairing in a $d$-wave state. It is straightforward to show that  it provides 
the superconducting orders in each pocket as
\begin{align}
\Da(\mathbf{k}) &= \Dp^\alpha\,(\coskx - \cosky)^2,
\label{eq:Dalpha}\\
\Db(\mathbf{k}) &=   -\Dp^\beta\,(\coskx - \cosky)^2,
\label{eq:Dbeta}\\
\Dg(\mathbf{k}) &=   \Dp^\gamma,
\label{eq:Dgamma}
\end{align}
where $\Dp^\gamma >\Dp^\beta >\Dp^\alpha$ based on the weight  of $\dz$ orbital in each pocket. 

The intralayer inter-orbital exchange $\Jxz$ provides a spin singlet inter-orbital pairing:  $\langle d_{z^2\uparrow}(\bm{k}) d_{x^2-y^2\downarrow}(-\bm{k})-d_{z^2\downarrow}(\bm{k})d_{x^2-y^2\uparrow}(-\bm{k}) \rangle=\Ds f(\bm{k})$, where $f(\bm{k})$ is the pairing form factor in the momentum space.  Similar to  $\Jperp$, it generates a sign change between pockets being of orbital in-phase (O$+$) and out-of-phase (O$-$).  In the pocket basis: O$-$ pockets ($\beta$,
$\gamma$) acquire the same gap sign, while the O$+$ pocket ($\alpha$) acquires an opposite gap sign. Moreover,  we need to specifically address the form factor in momentum space $f(\mathbf k)$ .  Considering the $B_{1g}$ symmetry of $\dx$, the total symmetry
of the inter-orbital pairing order parameter is:
\begin{equation}
\Gamma_\text{total} = \underbrace{\Gamma_{\dz}}_{A_{1g}}
\otimes \underbrace{\Gamma_{\dx}}_{B_{1g}} \otimes \Gamma_f,
\end{equation}
which requires $\Gamma_f = B_{1g}$ for an $s$-wave (A$_{1g}$) total symmetry and
$\Gamma_f = A_{1g}$ for $d$-wave (B$_{1g}$) total symmetry:
\begin{align}
f_s(\mathbf{k}) &= \coskx - \cosky \quad \text{[$s$-wave, }B_{1g}\text{]},
\label{eq:fs}\\
f_d(\mathbf{k}) &= \coskx + \cosky \quad \text{[$d$-wave, }A_{1g}\text{]}.
\label{eq:fd}
\end{align}
These form factors must be further projected onto the band eigenstates through the
orbital mixing angle $\sin 2\theta \propto (\coskx - \cosky)/\Delta_\pm$,
giving the band-space gap functions:
\begin{align}
\Delta^\text{band}_s(\mathbf{k}) &\propto (\coskx - \cosky)^2
\qquad [A_{1g},\; \text{s-wave}],
\label{eq:sband}\\
\Delta^\text{band}_d(\mathbf{k}) &\propto  \cos^2\kx - \cos^2\ky
\qquad [B_{1g},\;\text{d-wave}].
\label{eq:dband}
\end{align}
The $s$-wave channel selects by a large margin over the $d$-wave channel. The d-wave
band gap $\cos^2\kx - \cos^2\ky$ vanishes on both the zone diagonal ($k_x = k_y$)
and the zone axes ($k_x = 0$ or $k_y = 0$), and its overlap with all three
pockets is extremely small. Namely, the net overlap of the $d$-wave form
factor is negligible on all pockets. By contrast, the $s$-wave band gap
$(\coskx - \cosky)^2$ is fully symmetric (A$_{1g}$) and its overlap is
\textit{the largest on the $\beta$ pocket} for two reasons: (i) $\beta$ sits near
$M$ where $(\coskx - \cosky)$ is large; and (ii) $\beta$ has the strongest
$\dz$--$\dx$ hybridization among all pockets.The collaborative Fermi
surface rule therefore unambiguously selects the s-wave inter-orbital channel.

Combining both AFM channels, with $\Jperp$ assigning $\pm\Dp$ by mirror parity
and $\Jxz$ assigning $\pm\Ds(\coskx-\cosky)^2$ by orbital parity.  We obtain the general gap on each pocket as:
\begin{align}
\Da(\mathbf{k}) &= +(\Dp^\alpha +\Ds^\gamma)\,(\coskx - \cosky)^2,
\label{eq:Dalpha}\\
\Db(\mathbf{k}) &=-(\Dp^\beta+\Ds^\beta)\,(\coskx - \cosky)^2,
\label{eq:Dbeta}\\
\Dg(\mathbf{k}) &= +\Dp^\gamma -\Ds^\gamma\,(\coskx - \cosky)^2,
\label{eq:Dgamma}
\end{align}
where $\Dp^i\propto\Dp, i=\alpha,\beta,\gamma$ have the same sign so for  $\Ds^i\propto\Ds, i=\alpha,\beta,\gamma$.  In practice, we can ignore the inter-oribtal  pairing contribution on $\gamma$ as $(\coskx-\cosky)^2 \approx 0$ near $M$  is small.

The decisive role of the sign between $\Dp$ and $\Ds$ is now transparent.  The pairings in both channels can enhance  the superconducting gap on $\alpha$ and $\beta$ pockets in the $s$-wave state. The enhancement also requires  $\Dp$ and $\Ds$ must have the same sign. Namely, when
$\Dp$ and $\Ds$ carry the same sign--- the energetically preferred
state --- the two contributions to $\alpha$ and $\beta$ add constructively on
both pockets. In particular on  the $\beta$ pocket,  both $\Dp^\beta$ and $\Ds^\beta$ are relatively large and the enhancement is \textit{large} around the zone boundary where $(\coskx-\cosky)^2$ is maximum(namely, the anti-nodal region like the d-wave in cuprates). 

The above strong-correlation analysis can be compared with intermediate-coupling functional
renormalization group (FRG) calculations~\cite{Wang327prb,XWu,zhan2024cooperation,WangQH2025_prl,LeCC2025,ZhanJ_nonlocalV}, which
approach the problem from the opposite limit and have also found indications of a
$s^\pm$-state. In particular, the sign changes in momentum space of both $s^\pm$-states are almost identical. The agreement between the two different approaches strengthens confidence in
the physical picture. In FRG calculations, we can find that the critical magnetic fluctuation are given by the momentum vector $q_4$ as shown Fig.\ref{fig2}, connecting the $\alpha$ pocket to $\beta$ pocket~\cite{Wang327prb,XWu}. We have shown here that  $\alpha$  and $\beta$ have opposite characters in terms of both mirror eigen values and orbital relative phase. Therefore, this  scattering  drives antiferromagnetic fluctuations between layers as well as between orbitals. Within this spirit, it is not surprising  that the strong-correlation analysis and weak-interaction calculation converges to $s^\pm$ state. 

However, there is an important difference. In the intermediate-coupling FRG approach,
a significant on-site same-orbital pairing appears in addition to the inter-orbital
and interlayer channels. This on-site same-orbital pairing is intra-orbital and
momentum-independent, and it also carries the opposite sign from the interlayer
pairing. Such on-site same-orbital pairing is \textit{not allowed} in the
strong-correlation limit, where the Mott constraint (one $\dz$ electron per site,
$U_{zz} \to \infty$) strictly forbids double occupancy of the same orbital on the
same site.

We can argue that the on-site same-orbital pairing appearing in the weak-coupling
calculation is the effective manifestation of the inter-orbital pairing derived
here. In the weak-coupling approach, the orbital degree of freedom is integrated
out and the pairing is projected onto an effective band description. The
inter-orbital singlet bond between $\dz$ and $\dx$ on NN sites, when projected onto
the $\dz$ sector by tracing over the $\dx$ degree of freedom, generates an effective
on-site $\dz$ pairing amplitude with the same sign structure. The on-site character
reflects the short-range (NN bond) nature of the inter-orbital exchange $\Jxz$: in
the long-wavelength limit, the NN inter-orbital pairing at momentum $\mathbf{k}$ has
a large $\mathbf{k}=0$ (on-site) component. Therefore, the on-site pairing in the
weak-coupling approach and the inter-orbital pairing in the strong-coupling approach
are two complementary descriptions of the same physical process, linked by the
orbital hybridization that connects $\dz$ and $\dx$ through $t_{xz}$.

\section{Discussion}

{\it Comparison with cuprates and iron-based superconductors:}
Ni327 satisfies the gene requirements for high-$\Tc$ superconductivity in a
qualitatively new way. The quasi-2D bilayer NiO$_2$ structure isolates both $\dz$
and $\dx$ near the Fermi energy; the near-half-filled $\dz$ orbital provides the
strongly correlated sector that drives superexchange; and anion-mediated exchange
through both the apical and in-plane oxygens provides the pairing force. What
distinguishes Ni327 from all known families is the strong hybridization between two $e_g$ orbitals. The presence of both orbitals lays out the \textit{inter-orbital} nature
of the dominant exchange $\Jxz$: the half-filled orbital ($\dz$) and the
exchange-pathway orbital ($\dx$) are different, and the $B_{1g}$ symmetry of $\dx$
imprints a non-trivial momentum structure on the pairing. Table~\ref{tab:comparison}
summarizes the key similarities and differences among the three families.
\begin{table*}[t]
\centering
\caption{Comparison of the gene principle realization in cuprates, iron-based
superconductors, and Ni327. NN = nearest neighbor, NNN = next-nearest neighbor.}
\label{tab:comparison}
\begin{ruledtabular}
\begin{tabular}{lccc}
 & Cuprates & Iron pnictides & Ni327 (this work) \\
\hline
Active orbital(s) & $\dx$ (single) & $d_{xy}$-type (multi) & $\dz$ + $\dx$ (two) \\
Half-filled orbital & $\dx$ & $d_{xy}$ & $\dz$ \\
Dominant exchange & NN $\dx$--O--$\dx$ ($J_1$) & NNN Fe--As/Se--Fe ($J_2$) & Interlayer $\dz$--O$_\text{ap}$--$\dz$ ($\Jperp$) \\
                  &  &  & + intralayer $\dz$--O$_\text{in}$--$\dx$ ($\Jxz$) \\
Exchange character & Intra-orbital, in-plane & Intra-orbital, in-plane & In-plane inter-orbital + interlayer \\
Pairing form factor & $\coskx - \cosky$ & $\coskx\cosky$ & $(\coskx-\cosky)^2+\delta$ on $\beta$ \\
Pairing symmetry & d-wave & Extended s-wave ($\spm$) &  anisotropic s-wave s($^\pm$)  \\
Sign change location & Between $k_x$ and $k_y$ bonds & Between $\Gamma$ and $M$ pockets & Between mirror-even/odd pockets  \\
Gap structure & Nodal & Fully gapped & Fully gapped, anisotropic on $\beta$ \\
Dominant pocket & Large hole FS &  $M$ electron FS & $\beta$ large hole (mirror-odd, strongly hybridized) \\
\end{tabular}
\end{ruledtabular}
\end{table*}

In our analysis, it is clear that the $\beta$ pocket is the most important pocket in driving superconductivity where both AFM exchange couplings can collaboratively contribute  the superconducting pairing. If the pairing is purely attributed to  them, we would expect nodes along the diagonal direction in momentum space.  However, since it is $s$-wave,  there is not only no symmetry protection for the nodes, but also other mechanism can produce additional pairing channels to destroy the nodes. For example, by proximity effect, we should also expect the existence of  interlayer pairing between $\dx$ orbitals. In such case, we would expect the gap function on $\beta$ pocket as $\Delta_0+\Delta_1(cosk_x-cosk_y)^2$. Therefore, the anisotropy in $\beta$ pocket is a direct fingerprint of the
$(\coskx-\cosky)^2$ form factor and should be measurable by angle-resolved
spectroscopy or STM gap maps on Ni327 thin films.

Our analysis suggests that the presence of $\gamma$ pocket is not a necessary condition to obtain high $T_c$.   Since $\gamma$ pocket is mirror-even but orbital out-of-phase, from the strong-correlation view, the presence of $\gamma$ can strengthen the interlayer pairing but weaken the inter-orbital pairing. This physics is also captured in the FRG calculation as shown in Ref.~ \cite{LeCC2025,ZhanJ_nonlocalV}. 

Our approach can naturally extend to predict the superconducting state in trilayer nickelates Ni4310~\cite{Kuroki_trilayer,Dagotto_trilayer,WangQH_trilayer_FRG,YangF_trilayer,WuCJ_trilayer} in which the electronic structure has  an additional $\beta'$  pocket on Fermi surfaces besides  the three $\alpha, \beta$ and $\gamma$ pockets. With the NN inter-layer
AFM and the NN intralayer  inter-orbital AFM acting as the pairing forces, the predicted superconducting state on the trilayer nickelates   is also the $s^\pm$ state. In this case, the sign on $\beta$ and $\beta'$ is opposite to $\alpha$ and $\gamma$. Moreover,  there is no orbital component from the middle layer on $\beta'$ pocket.  Therefore,  on the $\beta'$ pocket,  the superconducting pairing is purely contributed from the inter-orbital $J_{xz}$. We can predict that the gap size on $\beta'$ is smaller than $\beta$  and must be close to a pure $(cosk_x -cosk_y)^2$ momentum dependence form.

The sign change on momentum space can result in distinguished quasiparticle interference.
The mirror-selective QPI pattern predicted for the $s^\pm$
state~\cite{ZhangQPI2025} provides a powerful momentum-space probe of the sign
structure. Mirror-odd impurities (such as the inner apical oxygen vacancy, which
couples preferentially to $\dz$ and therefore acts as a mirror-odd scatterer)
connect the mirror-odd $\beta$ pocket to the mirror-even $\alpha$ and $\gamma$
pockets. Since these pockets carry opposite gap signs in the  $s^\pm$ state, such
scattering is enhanced in the superconducting state relative to the normal state.
Mirror-even impurities, by contrast, scatter only within the mirror-even sector
and show suppressed inter-pocket QPI. The distinct enhancement and suppression
pattern at specific scattering wavevectors $\mathbf{Q}_1$ (connecting $\alpha$
and $\beta$) and $\mathbf{Q}_2$ (connecting $\gamma$ and $\beta$) constitutes
a definitive momentum-space test of the $s^\pm$ sign structure, qualitatively
distinct from the QPI patterns expected for any other proposed pairing state
in Ni327.

\section{Conclusion}

In conclusion, we have identified the pairing symmetry of bilayer nickelate La$_3$Ni$_2$O$_7$
under high pressure as a $s^\pm$ state, determined by two leading AFM
exchange interactions: the interlayer $\dz$--$\dz$ exchange $\Jperp$ through the
apical oxygen (c-axis bond) and the intralayer inter-orbital $\dz$--$\dx$ exchange
$\Jxz$ through the in-plane oxygen (ab-plane NN bond). 

The pairing symmetry is determined by the collaborative Fermi surface rule: because
$\dx$ carries $B_{1g}$ symmetry, the $s$-wave inter-orbital pairing has the $B_{1g}$
form factor $(\coskx - \cosky)$, which in band space generates $(\coskx - \cosky)^2$
with large overlap on the strongly hybridized $\beta$ pocket. The two channels contribute constructively to the large anisotropic gap on
$\beta$.

Ni327 represents a genuinely new class of high-$\Tc$ superconductor within the gene
principle framework---the first material where both the intra-orbital and inter-orbital pairing exchange  plays an important role in determining the $s^\pm$  state with 
distinctive experimental signatures in gap magnitude hierarchy.

\begin{acknowledgments}
We acknowledge the support by the National Natural Science Foundation of China (Grant No. NSFC-12494594, 12574151, 12447103 and 12447101), the Ministry of Science and Technology  (Grant No. 2022YFA1403900, No.2023YFA1407300) and  the New Cornerstone Investigator Program.
\end{acknowledgments}


\bibliography{references_new250806}
	\bibliographystyle{apsrev4-1}

\end{document}